\documentclass[aps,prd,eqsecnum,showpacs,amsmath,amssymb,superscriptaddress,preprint,nofootinbib]{revtex4-1}

\usepackage{amssymb,latexsym,amsmath,enumerate,amsthm}
\usepackage{graphicx}
\usepackage{color}
\usepackage{tikz}           
\usepackage{pgfplots}

\begin{document}
	
\title{Collapsing Wormholes Sustained by Dustlike Matter}%

\author{Pavel E. Kashargin}
\email{pkashargin@mail.ru}

\author{Sergey V. Sushkov}
\email{sergey$_\,$sushkov@mail.ru}

\affiliation{Institute of Physics, Kazan Federal University, Kremliovskaya str. 16a, Kazan 420008, Russia}

\begin{abstract} 
	{It is well known that static wormhole configurations in general relativity (GR) are possible only if matter threading the wormhole throat is ``exotic''---i.e., violates a number of energy conditions. For this reason, it is impossible to construct {static} wormholes supported only by dust-like matter which satisfies all usual energy conditions. However, this is not the case for non-static configurations. 
    In 1934, Tolman found a general solution describing the evolution of a spherical dust shell in GR. In this particular case, Tolman's solution describes the collapsing dust ball; the inner space--time structure of the ball corresponds to the Friedmann universe filled by a dust.    
  	In the present work we use the general Tolman's solution in order to construct a dynamic spherically symmetric wormhole solution in GR with dust-like matter. 
  	The solution constructed represents the collapsing dust ball with the inner wormhole space--time structure. It is worth noting that, with the dust-like matter, the ball is made of satisfies the usual energy conditions and cannot prevent the collapse. We discuss in detail the properties of the collapsing dust wormhole.}
\end{abstract}

\maketitle
 \section{Introduction}

Wormholes are solutions of general relativity which possess a throat. 
We define a wormhole throat as a closed 2-dimensional hypersurface of minimal area. 
Such solutions were first mentioned in the works~\cite{Flamm, EnstRos, Wheeler1, Wheeler2}, but the greatest interest in these objects arose after the work of Michael Morris and Kip Thorne in 1988~\cite{MorTho}. 
As noted by the authors, one of the ``exotic'' properties of these objects is that the presence of a throat leads to a violation of the energy conditions for the energy-momentum tensor of matter.
An overview of research on wormholes can be found, for example, in~\cite{Visser, LoboRew}.
Wormholes have been considered in various theories of gravity, including the Brans-Dicke theory of gravity~\cite{Agnese}, in the Einstein-Born-Infeld theory of gravity~\cite{Arellano}, in the Einstein-Gauss-Bonnet theory~\cite{Bhawal}, in $f(R)$ gravity~\cite{Miguel}, in Rastall's theory of gravity~\cite{Halder} and many other theories.
Various models of matter were considered as a source of wormholes: 
scalar and electromagnetic fields~\cite{Arellano, Bronnikov1, Bronnikov2}, Chaplygin gas~\cite{ChaplyginLobo}, various models of phantom energy and quintessence, including models of matter with the structure of the energy--momentum tensor of a perfect fluid~{\cite{Sushkov1, Kuhfittig2, Lobo3, Kuhfittig:2016, Sahoo}}.
Models of wormholes were also built using the thin-shell formalism, the~works of~\cite{Visser2, Visser3} being among the first.

Most of the literature is devoted to the study of static and spherically symmetric wormholes. 
An~important generalization of these studies is the study of dynamic solutions.
Dynamic wormholes have been considered in various aspects. One of the ways for building dynamic models ~\cite{Kar, KarSahdev, Kim, Roman, Kuhfittig3} is to add a time-dependent scaling factor to the metric.
Dynamic models of wormholes were constructed using the thin shell formalism~\cite{Anzhong}. 
General properties of an arbitrary dynamic wormhole are considered in the papers~\cite{Hayward2, Dynwh1}. 
In~recent papers~{\cite{Simpson, SimpsonLobo,SimpsonVisser}}, dynamic solutions have been constructed on the basis of a family of solutions describing regular black holes.  

It is worth noting that the throat of a wormhole is defined differently by different authors. 
In~the static case, the~different definitions agree with each other; in the dynamic case contradictions may arise. 
The~general defenition of the throat, including for time-dependent metrics, is considered, for example, in the following 
works~\cite{Dynwh1, TomikawaIzumi, Bittencourt}. We will use the definition that was used in the works~\cite{Kar, Kim, Roman}.

The~aim of this work is to construct a solution describing a dynamic wormhole in the theory of gravity with dust-like matter.
The~general solution of Einstein's equations in the theory of gravity with dust-like matter for a spherically symmetric metric was obtained by 
Tolman in 1934 ~\cite{Tolman34, Lemaitre33, Bondi47, Land, CosimoBambi}; the~solution will be summarized in Section~\ref{sec2}. 
This solution contains three arbitrary functions. 
In~Section~\ref{sec3} we construct a solution that describes a collapsing wormhole, choosing these functions in a certain way.
In~Section~\ref{sec4} we investigate the properties of the obtained solution.

\section{Gravitational Collapse of a Dust-Like Spherical Shell}
\label{sec2}
The spherically symmetric solution of the Einstein equations in the theory of gravity with dust-like matter was obtained by Tolman in 1934 \cite{Tolman34, Lemaitre33, Bondi47, Land, CosimoBambi}.  Briefly, the solution will be presented in this section.
Dust-like matter allows the choice of a frame that is both synchronous and co-moving. 
In this case, the spherically symmetric metric has the form\footnote{Henceforth, we denote the speed of light as 
	$c =1$.}: 
\begin{equation}\label{metric-4}
ds^2=d\tau^2-e^{\lambda(\tau,R)}dR^2-r^2(\tau,R)\big[d\theta^2+\sin^2\theta d\varphi^2\big],
\end{equation}
where $\lambda(\tau,r)$ and $r(\tau,R)$ are functions of $\tau$ and $R$. 
The energy--momentum tensor of dust-like matter in the co-moving frame has the form
\begin{equation}
T_{i}^{j}=\varepsilon u_i u^j, \label{fluid}
\end{equation}
where $(u^{i})=(1,0,0,0)$ is a velocity four vector, and $\varepsilon$ is the energy density. 
Below, we present the solution of the Einstein equations for the metric (\ref{metric-4}) and the energy--momentum tensor (\ref{fluid}). 
The function $\lambda(\tau,R)$ has the form
\begin{equation}
e^{\lambda(\tau,R)}=\dfrac{r'^2}{1+f(R)},\label{E5}
\end{equation}
where the prime means differentiation with respect to the coordinate $R$, the dot means differentiation with respect to the coordinate $\tau$,  $f(R)$ is an arbitrary function, and $f(R)$ satisfies the condition $1+f>0$. Function $r(\tau,R)$  can be represented in parametric form. 
In case $f>0$, the function has the form:
\begin{eqnarray}
&r=\dfrac{F(R)}{2f(R)}\left[ \cosh\eta-1 \right],
\quad
{\tau_0}(R)-\tau=\dfrac{F(R)}{2f(R)^{3/2}}\left[ \sinh\eta -\eta\right];&\label{Tolman1}
\end{eqnarray}
in the case of $f<0$, the function has the form:
\begin{eqnarray}
&r=\dfrac{F(R)}{2|f(R)|}\left[1- \cos\eta \right],\quad
{\tau_0}(R)-\tau=\dfrac{F(R)}{2|f(R)|^{3/2}}\left[\eta- \sin\eta\right],\label{Tolman2}& 
\end{eqnarray}
where $F(R)$ and ${\tau_0}(R)$ are an arbitrary functions. 
In the case of $f=0$, the function has the form:
\begin{eqnarray}
r=\left(\dfrac{9F}{4}\right)^{1/3}\left[{\tau_0}(R)-\tau\right]^{2/3}.
\end{eqnarray}

The energy density $\varepsilon$ is:
\begin{equation}
8\pi k \varepsilon=\dfrac{F'}{r'r^2 }.\label{varepsilon}
\end{equation}

The scalar curvature $\mathcal{R}$ and the Kretschman scalar $\mathcal{K}$ for the Tolman solution (\ref{metric-4}--\ref{Tolman2}) read
\begin{eqnarray}
& \mathcal{R} = \dfrac{F'}{r^2r'}, \label{R} &\\
& \mathcal{K} = \dfrac{3F'^2}{r^4r'^2}+ \dfrac{16 F^2}{r^6} - \dfrac{8FF'}{r^5r'}. \label{K}&
\end{eqnarray}

The scalar curvature and the Kretschmann scalar diverge at $r=0$ or $r'=0$ (in the case $F'\neq 0$), the solution in this case has a singularity.

The general solution (\ref{metric-4})--(\ref{Tolman2}) 
depend on three arbitrary functions $f(R)$, $F(R)$ and ${\tau_0}(R)$.  
The solution allows an arbitrary transformation of the radial coordinate $\tilde{R}=\tilde{R}(R)$, where $\tilde{R}(R)$ is a differentiable function. 
Using this, we can give any of the three functions $f(R)$, $F(R)$ or ${\tau_0}(R)$ some specific form. 
Thus, only two of the three functions can be considered arbitrary.
The particular cases of the solution are the Schwarzschild solution and the Friedmann model with dust.

\section{Throat Conditions}
\label{sec3}
In this section, we will obtain the conditions under which the general solution (\ref{metric-4})--(\ref{Tolman2}) will describe the wormhole geometry.
A characteristic feature of wormholes is the presence of a throat---i.e., the presence of a space-like closed two-dimensional surface of the minimum area.
Various definitions of the mouth of a wormhole are considered in the literature \cite{Dynwh1, TomikawaIzumi, Bittencourt, Kar, Kim, Roman}, 
we will follow the approach used, for example, in works \cite{Kar, Kim, Roman}.

In order to determine the throat conditions, following the approach used in \cite{Roman}, we constructed an embedding diagram for the metric (\ref{metric-4}). 
For convenience, we rewrote the function $f(R)$ in the form $f(R)=-b(R)/R$,  where $ b (R) $ is an arbitrary function. 
Taking into account the relation (\ref{E5}), the metric (\ref{metric-4}) can be represented as
\begin{equation}\label{metric-41}
ds^2=d\tau^2-\dfrac{r'^2 dR^2}{1-b/R}-r^2\big[d\theta^2+\sin^2\theta d\varphi^2\big].
\end{equation}

\begin{figure}[h]
\begin{center}\begin{tabular}{ccc}
\pgfplotsset{width=7cm,height=6.8cm}
\begin{tikzpicture}
\begin{axis}[
	axis y line = left,
    axis x line = middle,
		 xlabel     = $\rho$,
	 xtick=\empty, ytick=\empty,
		samples     = 160,
    domain      = 0.5:3,
    xmin = 0.5, xmax = 3,
    ymin = -1.8, ymax = 1.8,
]
\addplot[blue,thick] { sqrt(x-1)};
\addplot[blue,thick] { -sqrt(x-1)};
 \coordinate [label=right:$z$] (A) at (0cm, 5cm);   
\end{axis}
\end{tikzpicture}
& \,\,\,\,\,\,\,\,\,\,&
\begin{tikzpicture}
\begin{axis}[view={60}{30},axis lines=none,colormap/cool, fill opacity=.9]
	\addplot3[surf,
		samples=30,
        domain=-2:2,y domain=0:2*pi,
        z buffer=sort]
        ({(x^2+1) * cos(deg(y))},
		 {(x^2+1) * sin(deg(y))},
		 x);
\end{axis}
\end{tikzpicture}
\end{tabular}\end{center}
\caption{The embedding diagram. The function $\rho(z)$ is shown on the left panel, and the surface obtained by rotating the curve $\rho(z)$ about the axis $Oz$ is shown on the right panel.\label{fig7} }
\end{figure}
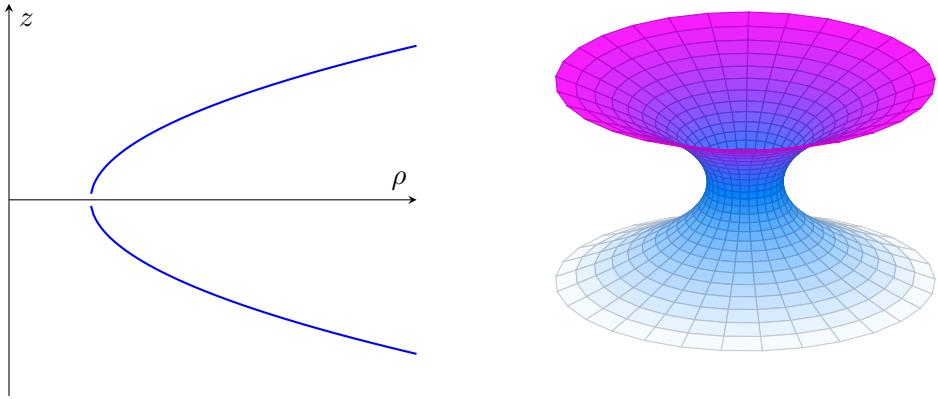

Metric (\ref{metric-41}) is defined for $1-b(R)/R>0$, and for $b(R)=R$ has a singularity.
Consider the metric (\ref{metric-41}) on the spatial slice $\tau=const$ and $\theta=\pi/2$
\begin{equation}\label{3_d}
dl^2=\dfrac{r'^2 dR^2}{1-b/R}+r^2d\varphi^2,
\end{equation}
as a metric on a surface of revolution $\rho=\rho(z)$ embedded in a three-dimensional space with an Euclidean metric
\begin{equation}\label{R3d}
dl^2=dz^2+d\rho^2+\rho^2d\varphi^2,
\end{equation}
where $z$, $\rho$ and $\varphi$ are cylindrical coordinates (see Figure \ref{fig7}).
Comparing (\ref{3_d}) and (\ref{R3d}), we get
\begin{eqnarray}
&\rho^2=r^2,&\\
&dz^2+d\rho^2=\dfrac{r'^2dR^2}{1-b/R}.&
\end{eqnarray}
Taking into account that for constant $\tau=const$
\begin{eqnarray}
&d\rho|_{\tau_0}=r'_R dR,&
\end{eqnarray}
we have
\begin{eqnarray}
&\dfrac{d\rho}{dz}=\left(\dfrac{R}{b}-1\right)^{1/2},\quad \dfrac{d^2\rho}{dz^2}=\dfrac{b-R b'}{2 r' b^2}\label{ee2}.&
\end{eqnarray}
%

The throat of the wormhole will have the shape of a sphere, which is located at a certain value of the radial coordinates $R=R_{th}$. 
On the embedding diagram, the sphere $R=R_{th}$ corresponds to a circle of radius $\rho$ on the surface of revolution; 
at the throat the radius of the circle $\rho(z)$ has a minimum.
Conditions for the minimum of the function $\rho(z)$ at $R=R_{th}$ have the form:
\begin{eqnarray}
&\dfrac{d\rho}{dz}\Big|_{R_{th}}=0,\quad \dfrac{d^2\rho}{dz^2}\Big|_{R_{th}}>0.&\label{ee4}
\end{eqnarray}
Comparing (\ref{ee2}) and (\ref{ee4}), we get {throat conditions for the metric} (\ref{metric-41}):
\begin{eqnarray}
&b(R_{th})=R_{th},\label{ee3}\\
&\dfrac{1-b'(R_{th})}{r'b(R_{th})}>0\label{ee5}.&
\end{eqnarray}

We will assume that the radial coordinate takes the following positive range of values: 
$0<R_{th}<R<+\infty$, where the value $R=R_{th}$ corresponds to the throat. 
The condition (\ref{ee3}) implies that the function $b(R)$ is positive at the throat: $b(R_{th})>0$.  
Therefore,  $b(R)$ is positive in some neighborhood of the throat by continuity. 
Therefore, the function $f=-b/R$ is negative in the vicinity of the throat, 
and to construct a wormhole model  we will consider the solution (\ref{Tolman2}) with a negative value of $f$.

\section{Collapsing Wormhole Model with Dust-like Matter}
\label{sec4}
The general solution (\ref{metric-4})--(\ref{Tolman2}) of the Einstein equations for a spherically symmetric metric in the theory of gravity with dust-like matter depend on three arbitrary functions $f(R)$, $F(R)$ and ${\tau_0}(R)$. In this section, we will consider a special case describing a wormhole. 
We choose arbitrary functions as follows\footnote{This case is not the only one possible. Further we will restrict ourselves to considering this example as one of the simplest.}: 
\begin{eqnarray}
&F=R,\quad {\tau_0}(R)={\tau_0}, \quad f(R)=-\dfrac{b}{R},\label{FT}&
\end{eqnarray}
where ${\tau_0}$, $b$ are constants, $b>0$ and $f<0$. The solution takes the form:
\begin{eqnarray}
&ds^2=d\tau^2-\dfrac{r'^2dR^2}{1-b/R}-r^2(t,R)\big[d\theta^2+\sin^2\theta d\varphi^2\big],\label{Twh1}&\\
&r=\dfrac{R^2}{2b}\left[1- \cos\eta \right],\quad {\tau_0}-\tau=\dfrac{R^{5/2}}{2b^{3/2}}\left[\eta- \sin\eta\right],\label{Twh3}&
\end{eqnarray}
where $R\in[b,+\infty)$. In this case, the first throat condition (\ref {ee3}) is satisfied for $R=b$. 
The second condition (\ref{ee5}) will be satisfied when $\left.r'\right|_{R = b}>0$. 
We calculate the derivative of the function $r(\tau,R)$ (\ref{Twh3}):
\begin{eqnarray}
&r'=\dfrac{R}{b(1-\cos\eta)}\left[
(1-\cos\eta)^2-\dfrac{5}{4}\sin\eta(\eta-\sin\eta) \right].\label{r'}
\end{eqnarray}

For  $\eta\in(0,2\pi)$ and $R\in[b,+\infty)$ the expression  (\ref{r'})  is positive: $r'>0$. 
Therefore, for the solution (\ref{Twh1}, \ref{Twh3}), the throat conditions (\ref{ee3}, \ref{ee5}) are satisfied for $R=b$.

To build a model of a wormhole, two identical sets $M_{+}$ and $M_{-}$ are required, each of which has a metric (\ref{Twh1}, \ref{Twh3}) with the same parameter $b$ 
($b_{+}=b_{-}$):
$$
M_{\pm}=\left\{ (\tau,R,\theta,\varphi)| \; R\in [b,+\infty) \right\}.
$$
$M_{+}$ and  $M_{-}$ are manifolds with boundaries. The boundary is a timelike hypersurface
$$
\Sigma_{\pm}=\left\{ (\tau,R,\theta,\varphi)| \; R=b \right\}.
$$ 

We identify $M_{+}$ and $M_{-}$ by boundaries. We get a new manifold $M=M_{+}\cup M_{-}$.
The resulting space $M$ has a wormhole geometry that connects two space--times, in each of which the radial coordinate takes the values $R\in[b,+\infty)$.
The throat of the wormhole corresponds to the value of the radial coordinate $R=b$.

As $R\to b$, one of the metric coefficients in (\ref{Twh1}), approaches infinity. 
However, the values of $r'$ (\ref{r'}) and $r$ (\ref{Twh3}) are nonzero for $\eta\in(0,2\pi)$, and the scalar curvature (\ref{R}) and the Kretschmann scalar (\ref{K}) are regular in this case. Thus, the singularity of the metric at the throat has a purely coordinate character.

It is known \cite{Visser2, Visser3, Lanczos, Darmois, Lichnerovich, Israel, Mars}, 
that the energy--momentum tensor on the shell $S_{ij}$ is proportinal to the Dirac delta function:
$$
T_{ij}=T_{ij}^{+}+T_{ij}^{-}+\delta(R-b)S_{ij},
$$
where $T_{ij}^{\pm}$ is the energy--momentum tensor in the corresponding region $M_{\pm}$. 
$S_{ij}$ is calculated as follows: 
$$
S_{ij}=\dfrac{1}{8\pi G}\left( [k_{ij}]-[k]h_{ij} \right)
$$
where $k_{ij}$ is the second fundamental form of $\Sigma$, 
$[k_{ij}]$ is the difference of limiting values of the second fundamental form on both sides of the surface $\Sigma$ in $M^{\pm}$
$$
[k_{ij}]=k_{ij}|_+-k_{ij}|_-=\left[h_i^{\alpha}h_j^{\beta}\nabla_{\beta}n_{\alpha}\right],
$$
$h_{\alpha\beta}$ is a projection tensor on $\Sigma$,
$$
h_{\alpha\beta}=g_{\alpha\beta}-n_{\alpha}n_{\beta},
$$
and $n_{\alpha}$ is the normal vector to $\Sigma$. 
In our case, the unit normal vector to the surface has the form
$
(n^i)=\left( 0, \pm e^{-\lambda/2},0,0 \right).
$
The tensor $[k_{ij}]$ on the surface $R=b$ 
$$
[k_{ij}]=
\left(
\begin{array}{cccc}
0 & 0 & 0 & 0\\
0 & 0 & 0 & 0 \\
0 & 0 & 2r\sqrt{1-b/R} & 0 \\
0 & 0 & 0 & 2r\sqrt{1-b/R}\sin^2\theta 
\end{array}\right)
$$
equals zero. Consequently, the shell momentum energy tensor
$
S_{ij}=0
$
is zero. Thus, there is no material shell on the surface $\Sigma$, and the energy--momentum tensor of the model contains only dust-like matter.

\section{Analysis of the Model}

In this section, we will consider the dynamics of the model from the point of view of an external observer.
We will assume that the dusty matter is distributed inside some spherical layer.
Outside this layer, the geometry is described by the spherically symmetric Schwarzschild solution in the vacuum \cite{Land}:
\begin{eqnarray}
&ds^2=\left(1-\dfrac{r_g}{r}\right)dt^2-\left(1-\dfrac{r_g}{r}\right)^{-1}dr^2-r^2d\Omega^2,\label{Sh}&
\end{eqnarray}
where  $r_g=2M$ is a gravitational radius, $M$ is a mass that creates a gravitational field. 
In the absence of pressure, each element of matter moves along a geodesic path. All geodesics are radial due to spherical symmetry. 
From the point of view of an external observer, the radial coordinate of the surface of the star is a function of time $r=r(\tau)$, where $r(\tau)$ is the radial geodesic in the Schwarzschild space--time. 
It is convenient to represent the geodesic equation in parametric form \cite{Lightman}:
\begin{eqnarray}
&r=\dfrac{r_0}{2}(1+\cos\chi),\label{geod_Sh1}&\\
&\tau=\left(\dfrac{r_0^3}{8M}\right)^{1/2}(\chi+\sin\chi),\label{geod_Sh2}&\\
&u^t=\dfrac{dt}{d\tau}=\dfrac{(1-2M/r_0)^{1/2}}{1-2M/R},\label{geod_Sh3}&
\end{eqnarray}
where the parameter $\chi\in[0,\pi]$, $\tau$ is a time coordinate in the co-moving coordinate system, $r_0$ is the value of the radial coordinate at the initial moment of time $\tau=0$ ($\chi=0$). 
Expressions (\ref{geod_Sh1}--\ref{geod_Sh3}) are valid provided that at the initial moment of time $\tau=0$ the shell was at rest $dr/d\tau=0$.
Substituting expressions (\ref{geod_Sh1}--\ref{geod_Sh3}) into the metric (\ref{Sh}), we find the 3-geometry of the surface of the star in the external metric:
\begin{eqnarray}
&{}^{(3)}ds^2=d\tau^2-R^2(\tau)d\Omega^2&\label{ds_out}\\
&=\dfrac{r_0^3}{8M}(1+\cos\chi)^2d\chi^2-\dfrac{r_0^2}{4}(1+\cos\chi)^2d\Omega^2.\nonumber&
\end{eqnarray}

Now we will consider the internal geometry of the dusty layer (\ref{Twh1}),   (\ref{Twh3}). 
Let us assume that dust-like  matter is enclosed in a spherical layer $R\in[b,R_{0}]$, where $R_0$ is its outer radius.
The internal 3-geometry of the surface of the star is found by substituting expression (\ref{Twh3}) into the metric (\ref{Twh1}) at $R=R_0$:
\begin{eqnarray}
&{}^{(3)}ds^2=d\tau^2-r^2(R_0,t)d\Omega^2&\label{ds_in}\\
&=\dfrac{R_0^5}{4b^3}(1-\cos\eta)^2d\eta^2-\dfrac{R_0^4}{4b^2}(1-\cos\eta)^2d\Omega^2.\nonumber&
\end{eqnarray}

On the surface of the star the Schwarzschild metric should be smoothly matched with the intrinsic metric. 
Comparing (\ref{ds_out}) and (\ref{ds_in}), we see that these 3-geometries are smoothly sewn together if we identify the following parameters
\begin{eqnarray}
&\eta=\pi-\chi,\quad r_0=\dfrac{R_0^2}{b},\quad M=\dfrac{R_0}{2}. \label{junctioncons}&
\end{eqnarray}

At the initial moment of time $\tau=0$ ($\chi=0$), the dust-like layer with the outer radius $r_0$ begins to collapse from the rest position towards the center.
As follows from formulas (\ref{geod_Sh1}, \ref{geod_Sh2}), at the moment of collapse, the outer radius vanishes $r=0$, 
and the watch of the co-moving observer shows  $\tau_0=\dfrac{\pi r_0^{5/2}}{2b^{3/2}}$ ($\chi=\pi$). 
The time $\tau_0 $ is the lifetime of the wormhole. 
At the moment of collapse, solutions (\ref{Twh1}, \ref{Twh3}) are singular, since the scalar curvature and the Kretschmann scalar (\ref{R}, \ref{K}) approach infinity. 
By construction, $R_0>b$, which means that due to the relations (\ref{junctioncons}) the initial outer radius of the shell will be greater than the gravitational radius---i.e., $r_0>r_g$. From the point of view of an infinitely distant observer, gravitational collapse will last an infinitely long time.

\begin{figure}[h]
\centerline{	
\includegraphics[scale=0.6]{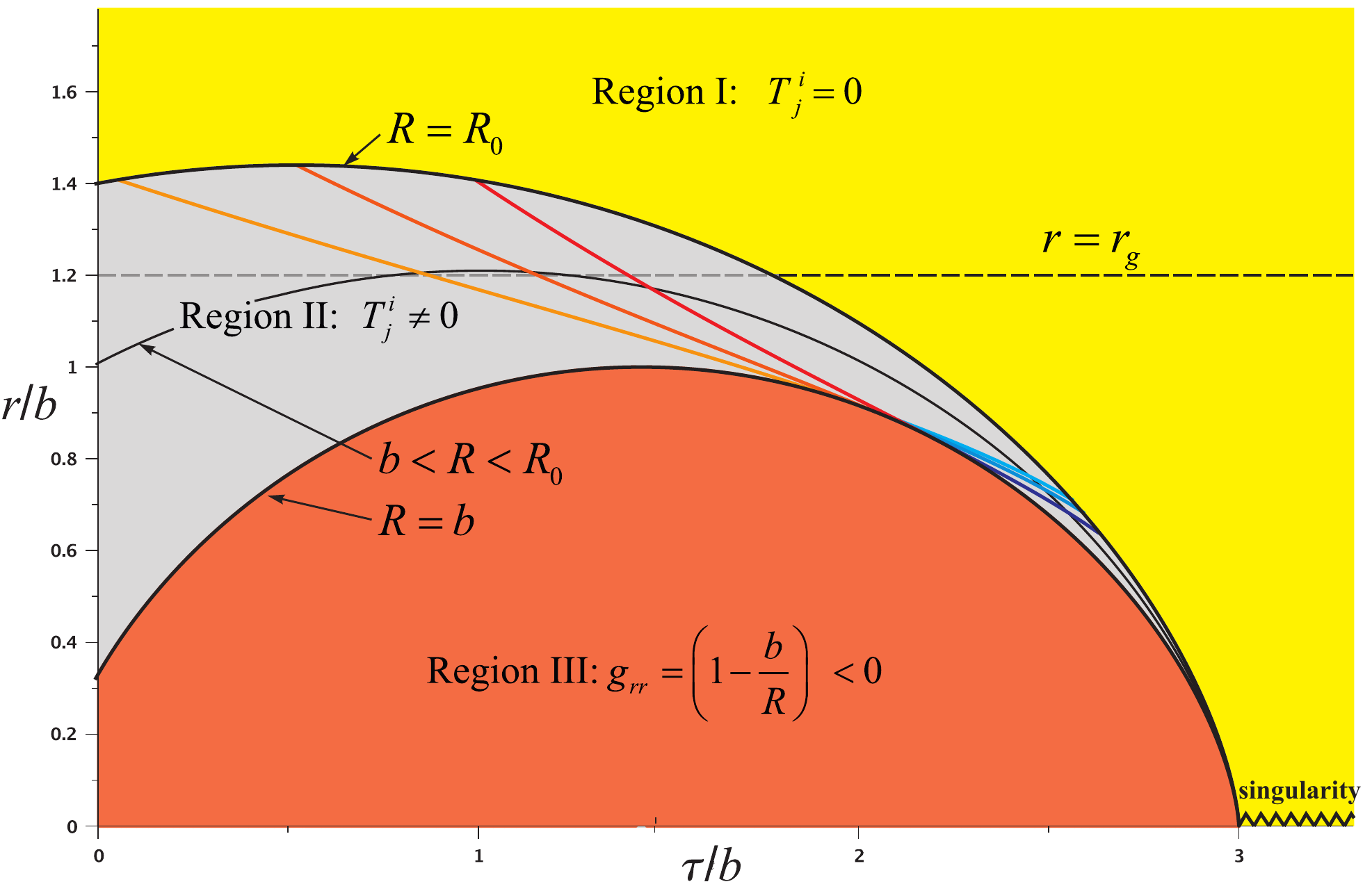}
}
\caption{The figure shows the graphs of the function $r(R,\tau)/b$ depending on $\tau/b$ for different values of $R$: $R=b$, $R=1.1b$ and $R=R_0=1.2b$.	
Red-blue lines are the null radial geodesics. (See the text for more details.)}
\label{fig2} 
\end{figure}

{
It is very important to answer the question: Is a photon able to cross the collapsing wormhole? To solve this problem we need to find null geodesics inside the collapsing dust ball. An equation for radial geodesics can be directly found from the condition $ds^2=0$. The metric (\ref{Twh1}) yields
\begin{equation}\label{geod_R}
\frac{dR}{d\tau}=\mp \frac{1}{r'}\,\sqrt{1-\frac{b}{R}},
\end{equation}   
where $r'$ is given by Eq. (\ref{r'}). Using the relations (\ref{Twh3}), one can rewrite Eq. (\ref{geod_R}) as follows
\begin{equation}\label{geod_r}
\frac{dr}{d\tau}=\left(\frac{b}{R}\right)^{1/2}\,
\left[-\frac{\sin\eta}{1-\cos\eta} \mp\sqrt{\frac{R}{b}-1}\right].
\end{equation}
Integrating this equation gives null geodesics in the coordinates $(r,\tau)$. Note that the sign $"-"$ corresponds to ingoing geodesics moving towards the wormhole throat, while $"+"$ corresponds to outgoing geodesics moving away from the throat. At the throat $R=b$ Eq. (\ref{geod_r}) reduces to
\begin{equation}
\left.\frac{dr}{d\tau}\right|_{R=b}=-\frac{\sin\eta}{1-\cos\eta},
\end{equation} 
therefore null geodesics are tangent to the cycloid $R=b$.

In the figure \ref{fig2} we illustrate the main features of the obtained solution. In particular, the dependence of the radius $r(R,t)$ of various layers of the dust-like shell $R\in[b,R_0]$ on time $\tau$ for $R_0=1.2b$ (see Eq. (\ref{Twh3})) is shown.
Region \textit{I}: $ R <b $ is not included in space--time and is not considered.
Region \textit{II}: $R\in[b,R_0]$ is filled with dust-like matter. 
Region \textit{III} is the empty area outside the dust-like shell.
The three solid curves show the dynamics of the dust layers over time: curve at $R=R_0$ is the outer shell; 
curve at $R=b$ is the wormhole throat; intermediate curve for the inner layer $b<R<R_0$, $R=1.1b$. 
The dash-dotted line corresponds to the gravitational radius $r=r_g$. 
In a finite proper time $\tau_0$, the shell reaches the gravitational radius and collapses towards the singularity $r=0$. 
Red-blue lines are the null radial geodesics. The red curves are ingoing geodesic lines of photons who started their travel towards the wormhole throat from the outer shell. These photons are reaching the throat in a finite time; at the point where the photon is crossing the throat, its geodesic line becomes tangent to the cycloid $R=b$. After crossing the throat, the photon continues its movement on the other side of the wormhole space--time. In the figure \ref{fig2} the blue curves correspond to outgoing geodesic lines of photons who started their travel towards the outer shell from the throat.  
}

%
\begin{figure}[h]
\centerline{
\includegraphics[scale=0.6]{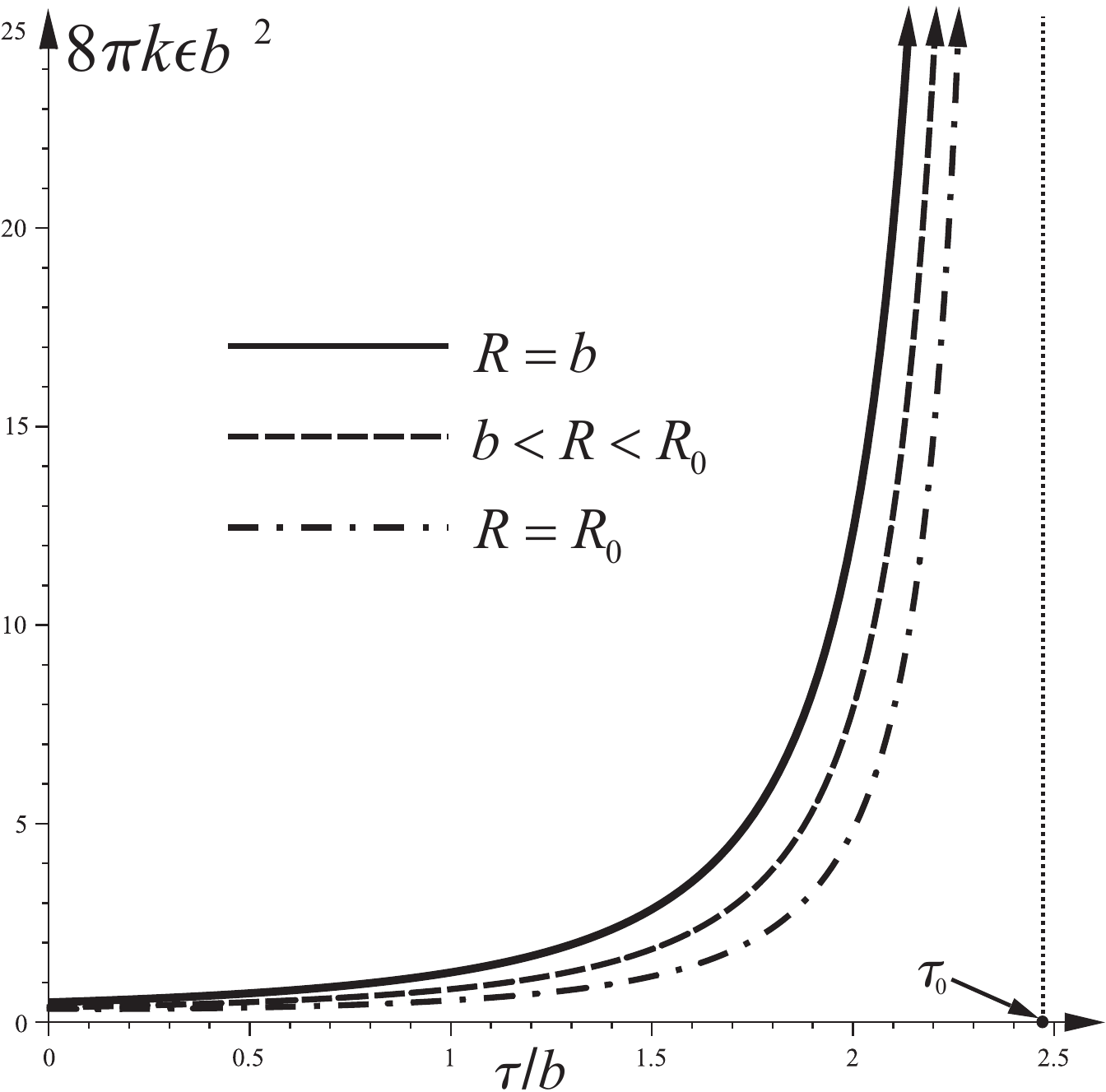}}
\caption{The figure shows the graphs of the energy density $8\pi k\varepsilon b^2$ depending on $\tau/b$ for different values of $R$: $R=b$, $R=1.1b$ and $R=R_0$, $R_0=1.2b$.  As $\tau\to\tau_0$ the energy tends to infinity, the vertical asymptote is shown by dash-dot on the right in the figure.}\label{fig3}
\end{figure}
Figure \ref{fig3} shows the dependence (\ref{varepsilon}) of the energy density $\varepsilon$ on the time $\tau$ for different values of $R\in[b,R_0]$.
The energy density is positive, and at the moment of collapse $\tau\to\tau_0$, the energy tends to infinity.
The energy--momentum tensor of ``ordinary'' matter satisfies certain physical requirements, one of which is the null energy condition:
$$
T_{\mu\nu}V^{\mu}V^{\nu}\leqslant 0
$$
for any null vector $V^{\mu}$. It is known \cite{Visser}, that the null energy condition can be violated in wormhole space--time.
Consider a region in the vicinity of the wormhole throat $R\in[b,R_0]$, where the model is described by solutions (\ref{Twh1}, \ref{Twh3}). 
We choose a null vector $V^{\ mu}=\left( 1,e^{-\lambda/2},0,0 \right)$ and calculate the following contraction:
\begin{equation}
T_{\mu\nu}V^{\mu}V^{\nu}=\varepsilon,\label{NEC}
\end{equation}
which is positive for $R\in[b,R_0]$, $\tau\in [0,\tau_0]$. Thus, in this model, the null energy condition is fulfilled.

We also present this metric in the limiting case $\tau\to\tau_0$ ($\eta\to+0$, $\chi\to\pi-0$):
\begin{eqnarray}
&
r(\tau,R)=R^{1/3}\left[\dfrac{3}{2}\left(\tau_0-\tau\right)\right]^{2/3},
&\\
&
ds^2=d\tau^2-\left(\dfrac{3}{2}\dfrac{\tau_0-\tau}{R}\right)^{4/3}\left[\dfrac{1}{12^{2/3}}\dfrac{dR^2}{1-b/R}+R^2\left(d\theta^2+\sin^2\theta d\varphi^2\right)\right]
&
\end{eqnarray}

\section{Conclusion}
It is well known that static wormhole configurations in general relativity (GR) are possible only if matter threading the wormhole throat is ``exotic''---i.e., violates a number of energy conditions. For this reason, it is impossible to construct {static} wormholes supported only by dust-like matter which satisfies all usual energy conditions. However, this is not the case for non-static configurations. 
In 1934, Tolman found a general solution describing an evolution of a spherical dust shell in GR. In this particular case, Tolman's solution describes the collapsing dust ball; the inner space--time structure of the ball corresponds to the Friedmann universe filled by a dust.    
In the present work we have used Tolman's solution in order to construct a dynamic spherically symmetric wormhole solution in GR with dust-like matter. 
For this aim, we have considered the specific subclass of arbitrary functions $f(R)$, $F(R)$ and $t_0(R)$ constituting Tolman's solution,
and written solutions (\ref{Twh1})--(\ref{Twh3}), which satisfy the throat conditions.
The solution constructed represents the collapsing dust ball with the inner wormhole space--time structure. 

%
%

To construct a wormhole model we considered two identical solutions $M_{+}$ and $M_{-}$ with metrics (\ref{Twh1}), (\ref{Twh3}), in each of which the radial coordinate $R$ takes the range of values $R\in[b,+\infty)$. The sets $M_{+}$ and $M_{-}$ are identified by their boundaries $\Sigma$. The result is a new manifold $M=M_{+}\cup M_{-}$ that has a wormhole geometry.
The throat of the wormhole corresponds to the junction surface $\Sigma$: $R=b$.
Due to the smoothness of the functions, the identity of $M_{+}$ and $M_{-}$ and to the throat conditions, the energy--momentum tensor $S_{ij}$ of the shell $\Sigma$ is equal to zero. Thus, the energy--momentum tensor of the constructed model contains only dust-like matter, and does not contain a contribution of the Dirac delta function on the junction surface.
For $\eta\in(0,2\pi)$, the scalar curvature (\ref{R}) and the Kretschmann scalar (\ref{K}) are regular, and the solution does not have singularities.
 
We assumed that the dusty matter is enclosed inside some spherical layer and considered the dynamics of a wormhole from the point of view of an external observer. 
Gluing solutions (\ref{Twh1}), (\ref{Twh3}) and the Schwarzschild solution, we obtained the condition (\ref{junctioncons}) that connects the model parameters, the gravitational radius and the mass of the dust ball.
The dynamics of dust layers over time is depicted in Figure \ref{fig2}.
At the initial moment of time, a dusty layer begins to collapse towards the center, at the moment of collapse the outer radius of the shell vanishes, and the clock of the co-moving observer shows $\tau_0={\pi r_0^{5/2}}/{2b^{3/2}}$. The $\tau_0$ is the wormhole's lifetime. 
From the point of view of an infinitely distant observer, gravitational collapse will last indefinitely.
At the moment of collapse, the scalar curvature and the Kretschmann scalar (\ref{R}), (\ref{K}) become singular and the solution has a singularity.
Figure \ref{fig3} shows the dependence of the energy density of the dust layer on time, the energy density is positive, and tends to infinity at the moment of collapse. It is shown that the null energy condition for this model is satisfied (\ref{NEC}). Therefore, the dust-like matter that the ball is made of satisfies the usual energy conditions and cannot prevent the collapse.

\acknowledgments{We acknowledge the contribution of Evdokim Isanaev on the very beginning of this work. P.E.K. and S.V.S. are supported by RSF grant No. 16-12-10401. Partially, this work was done in the framework of the Russian Government Program of Competitive Growth of the Kazan Federal University.}




\end{document}